\documentclass[aps,superscriptaddress,showpacs,twocolumn]{revtex4}

\usepackage[dvips]{graphicx}
\usepackage{amssymb}

\def \beq {\begin{equation}}
\def \eeq {\end{equation}}

\begin{document}
\title{Alternative numerical computation of one-sided L\'evy and Mittag-Leffler distributions}

\author{Alberto Saa}\email{asaa@ime.unicamp.br}
\address{Departamento de Matem\'atica Aplicada, UNICAMP, 13083-859, Campinas, SP, Brazil}
\author{Roberto Venegeroles}\email{roberto.venegeroles@ufabc.edu.br}
\address{Centro de Matem\'atica, Computa\c c\~ao e Cogni\c c\~ao, UFABC, 09210-170, Santo Andr\'e, SP, Brazil}

\date{\today}

\begin{abstract}
We consider here the recently proposed closed form formula in terms of the Meijer G-functions for the probability density functions $g_\alpha(x)$ of one-sided L\'evy stable distributions  with rational index $\alpha=l/k$,
  with  $0<\alpha<1$.
Since   one-sided L\'evy and Mittag-Leffler distributions are known to be related, this formula could also be useful for
calculating the probability density functions $\rho_\alpha(x)$ of
the
latter.
  We show, however,  that the formula is computationally inviable
  for fractions with large denominators, being unpractical even for some modest
  values of  $l$ and $k$. We present a fast and accurate numerical scheme,
  based on   an  early integral representation due to Mikusinski,
for the evaluation of  $g_\alpha(x)$ and $\rho_\alpha(x)$, their cumulative distribution
function and their derivatives
for any real index $\alpha\in (0,1)$. As an application, we explore some properties of these probability density functions. In particular, we determine the location and value  of their maxima
 as functions of the
  index $\alpha$. We show   that $\alpha \approx 0.567$ and
 $\alpha \approx 0.605$   correspond, respectively,
  to the   one-sided
L\'evy and Mittag-Leffler distributions with shortest maxima.
We close by discussing how
our results can elucidate some recently described dynamical behavior of
intermittent systems.
\end{abstract}

\pacs{  05.40.Fb, 02.50.Ng, 02.60.Jh}

\maketitle

\section{Introduction}

One-sided L\'evy stable distributions \cite{Levy1,Levy2} are ubiquitous in
many modern  research areas where quantitative and statistical analysis
play a major role. (For recent reviews, see, besides \cite{Levy2},
  the references of \cite{PG}.)
The  probability density function of one-sided L\'evy distribution of index $\alpha$, $g_\alpha(x)$,
 can be defined by means of its Laplace transform as \cite{Levy1}
\beq
\label{laplace}
\int_0^\infty e^{-sx}g_\alpha(x)\, dx= \exp\left(-s^\alpha\right),
\eeq
for $s\ge 0$, with $0<\alpha< 1$. Unfortunately, in spite of its broad applicability,
exact solutions of Eq. (\ref{laplace}) are
available only for a few particular values of $\alpha$. (See, for
 instance, the Appendix A of \cite{Barkai}. We notice also that there are
 some available Mathematica \cite{Math} and Matlab \cite{Matlab} packages for the numerical evaluation of $g_\alpha(x)$.) In this context,
the recent work of Penson and   G\'orska \cite{PG} is certainly
interesting and relevant since they describe a formal solution of Eq.
(\ref{laplace}) for any rational $\alpha$. In fact, they show that
a formula presented without proof in a table of inverse Laplace
transforms \cite{table} could be used to write
\beq
\label{meijer}
g_{l/k}(x) = \frac{\sqrt{kl}}{(2\pi)^{(k-l)/2}}\frac{1}{x}
G_{l,k}^{k,0}\left(
\frac{l^l}{k^k x^l}\left|
\begin{array}{c}
\Delta(l,0)\\
\Delta(k,0)
\end{array}
\right.
\right),
\eeq
where
$G_{p,q}^{m,n}\left(z\left|^{(a_p)}_{(b_q)}\right.\right)$ is the Meijer G-function \cite{math}
and $\Delta(k,a)$ is the list of $k$ elements given by
\beq
\Delta(k,a) = \frac{a}{k}, \frac{a+1}{k}, \cdots, \frac{a+k-1}{k}.
\eeq
We consider the formula (\ref{meijer}) an important advance.
Since the Meijer G-function is available in several
computer algebra systems, the function $g_{l/k}(x)$ could in principle be
evaluated with little programming effort. We notice that the restriction to
rational values of $\alpha$ in Eq. (\ref{meijer}) does not represent any
real
problem here. As we will see below, the function $g_\alpha(x)$ is
 continuous in $\alpha$ and, hence, one might compute from Eq. (\ref{meijer})
a rational $\alpha$ approximation for $g_\alpha(x)$  with any
prescribed accuracy.  Penson and   G\'orska \cite{PG} use
 Eq. (\ref{meijer}) to derive other series expression for
 $g_{l/k}(x)$ and to infer some of its properties. Certainly, the
 mathematical literature about the Meijer G-function (see, for instance,
 \cite{math} and the references therein) will be extremely
 valuable for the derivation of many other properties of $g_{l/k}(x)$
 defined by Eq. (\ref{meijer}).

Furthermore, since one-sided L\'evy and Mittag-Leffler distributions are known to be related \cite{Feller}, the formula (\ref{meijer}) is
 also relevant for
calculating the probability density functions $\rho^{(r)}_\alpha(x)$ of
Mittag-Leffler distributions with rational index $\alpha$. We recall that
$\rho^{(r)}_\alpha(x)$  is  also  defined from its Laplace transform as
well,
\beq
\label{laplace1}
\int_0^\infty e^{-sx}\rho^{(r)}_\alpha(x)\, dx=
\sum_{n=0}^{\infty}\frac{(-sr^{\alpha})^{n}}{\Gamma(1+n\alpha)},
\eeq
for $s\ge 0$, with $0<\alpha< 1$. The right-handed side of Eq. (\ref{laplace1})
is a particular case of the  so-called Mittag-Leffler function \cite{math}, which
reduces to the usual exponential   for $\alpha=1$.
The free parameter $r$ can
be fixed, for instance, by demanding a given   first moment for $\rho^{(r)}_\alpha(x)$.  In particular, since we have from Eq. (\ref{laplace1}) that
\beq
\label{trans}
 \rho^{(r)}_\alpha(q x) = q^{-1} \rho^{(r/q^{1/\alpha})}_\alpha(x),
\eeq
  for any $q > 0$, one can
assume $r=1$ without loss of generality. In this case, the superscript is
simply dropped.
  The
respective
cumulative distribution functions associated to $\rho_\alpha(x)$
and $g_\alpha(x)$ are known to be related by \cite{Feller}
\beq
\label{feller}
\Theta_\alpha(x) = 1 - \Lambda_\alpha\left({x^{-1/\alpha}}\right),
\eeq
which leads to
\beq
\label{feller1}
\rho_\alpha(x) = \frac{1}{\alpha }x^{-(1+1/\alpha)}g_\alpha\left({x^{-1/\alpha}}\right).
\eeq
The relation (\ref{feller1}) allows the computation of $\rho_\alpha(x)$ by means of the Meijer G-function for
rational $\alpha$, thanks to
the Penson and   G\'orska formula (\ref{meijer}). This is a considerable advance since, as in the previous case,
no closed form solution of Eq. (\ref{laplace1}) is known.

However, the condensed and apparently simple form of Eq. (\ref{meijer})
hides a practical pitfall. The evaluation of Eq. (\ref{meijer}) is
computationally viable only for modest values of $k$ and $l$. For
instance, by using the Maple procedure provided by Penson and   G\'orska \cite{procedure}, we
can plot  the graphics of $g_{2/3}(x)$
 for $x\in[0,2]$  instantaneously  in an Intel Core i7 computer
running  Maple version 14. In order to generate the same graphics
with, for instance, $l/k = 20/31$, some CPU minutes are necessary. For
$l/k = 200/301$, we need almost a half an hour to evaluate a single value of
$g_{l/k}(x)$! We could not evaluate Eq. (\ref{meijer}) for $l/k = 2000/3001$ in any reasonable amount of time.
Mathematica presents a similar performance.
These restrictions, obviously, jeopardize the practical utility
of expression (\ref{meijer}) since one cannot calculate in reasonable time
good approximations to the one-sided L\'evy   distribution
for any $\alpha$. One can understand the rapidity with the evaluation
of Eq. (\ref{meijer}) becomes unpractical when the
 values
of $l$ and $k$ increase by recalling the definition of the Meijer
G-function \cite{math}
\begin{eqnarray}
\label{def}
&& G_{p,q}^{m,n}\left(z\left|^{(a_p)}_{(b_q)}\right.\right) =  \\
&& \frac{1}{2\pi i}
\int_L
\frac{\prod_{j=1}^m\Gamma(b_j-s)\prod_{j=1}^n\Gamma(1-a_j+s)}
{\prod_{j=m+1}^q\Gamma(1-b_j+s)\prod_{j=n+1}^p\Gamma(a_j-s)}
z^s ds, \nonumber
\end{eqnarray}
where $L$ is a carefully chosen integration path on the complex plane.
It is possible also to write the Meijer G-function as
a sum of $m$ terms involving  $\Gamma$ function  products as those
ones of the integrand in Eq. (\ref{def})
and the generalized hypergeometric functions ${}_pF_{q-1}$ \cite{math}.  As one can see, when asking Maple to evaluate Eq. (\ref{meijer}) for
$l/k=2000/3001$, one is basically
 demanding the evaluation of an integral with
more than five thousands  $\Gamma$ function terms in the integrand, or an intricate combination of more than
three thousands generalized hypergeometric functions! Hence, it is not a surprise to have
a considerable performance degradation for large values of $l$ and $k$.
Another problem with the Maple procedure based in Eq. (\ref{meijer}) is that it does not deal efficiently
with reducible fractions. For instance, Maple is not able to
reduce
$g_{5/10}(x)$ to $g_{1/2}(x)$. Moreover, the  numerical evaluation of the
former is much more time and memory consuming than the latter.

The purpose of the present work is to show that one can compute
numerically, in an effective and efficient way,
 the probability density functions
$g_\alpha(x)$ and $\rho_\alpha(x)$
  with arbitrary real index
 $\alpha\in(0,1)$. Our start point is the
Mikusinski's  integral representation  for $g_\alpha(x)$ \cite{Mikusinski}
\beq
\label{Miku}
g_\alpha(x) =\frac{\alpha}{1-\alpha}  \frac{1}{\pi x}
\int_0^\pi ue^{-u }\,d\varphi,
\eeq
with $0<\alpha<1$,
where
\beq
\label{u}
u =  \frac{\sin(1-\alpha)\varphi}{\sin\varphi}\left(
\frac{\sin \alpha\varphi}{x\sin\varphi}
 \right)^{\alpha/(1-\alpha)}.
\eeq
The integral  representation (\ref{Miku})  has
already proven its relevance. In fact, Mikusinski used it
to derive more than 40 years ago the some very useful asymptotic expressions
for $g_\alpha(x)$, namely
\beq
\label{asymp1}
g_\alpha(x) \approx K\frac{\exp{\left(-Ax^{-\alpha/(1-\alpha)}\right)}}{x^{(2-\alpha)/(2-2\alpha)}},
\eeq
 valid for $x\to 0^+$ and
\beq
\label{asymp2}
g_\alpha(x) \approx M   x^{-(1+\alpha)},
\eeq
valid for  $x\to\infty$, where
 \begin{eqnarray}
 A &=& (1-\alpha)\alpha^{\alpha/(1-\alpha)}, \\
 K&=& \frac{\alpha^{1/(2-2\alpha)}}{\sqrt{2\pi (1-\alpha)}}, \\
 M &=& \frac{\sin\alpha\pi}{\pi }\Gamma(1+\alpha).
 \end{eqnarray}
It is easy to check from Eq. (\ref{Miku}) that $g_\alpha(x)$ is
non-negative and smooth in $\alpha\in(0,1)$. Due to Eq. (\ref{feller1}),
  and one has
the following asymptotic behavior for $\rho_\alpha(x)$
\beq
\label{asympML1}
\rho_\alpha(x) \to \frac{\sin\alpha\pi}{\alpha\pi }\Gamma(1+\alpha) ,
\eeq
  for $x\to 0^+$ and
\beq
\label{asympML2}
\rho_\alpha(x) \approx \frac{K}{\alpha} {x^{(2\alpha-1)/(2-2\alpha)}} {\exp{\left(-Ax^{1/(1-\alpha)}\right)}},
\eeq
valid for  $x\to\infty$.
In the next section, we show how to use Eq. (\ref{Miku}) to evaluate
numerically  one-sided L\'evy and Mittag-Leffler probability
 densities, their
cumulative distribution function,  and their
derivatives in a very efficient and reliable  way.

We notice that one-sided L\'evy stable distributions can be alternatively  expressed by means of Fox H-functions \cite{WGMN}, which are
 a further generalization of the Meijer G-functions
 (\ref{def}), and also by means of Wright
functions \cite{Mainardi}. Unfortunately, the current knowledge about the analytical structure of these functions is still little developed.
We wish also to stress here that the numerical computation of stable distributions is not a new problem and several algorithms are already available in the literature and even commercially. In particular, Nolan \cite{JPN} proposed a robust algorithm based
on the integration of the so-called Zolotarev's (M) representation for
stable distributions, which is the base of   Mathematica \cite{Math} and
Matlab \cite{Matlab} packages. An updated reference list on the subject
can be found in \cite{Ref}. However, as we will see, Mikusinski's  representation
(\ref{Miku}) allows the numerical evaluation of one-sided L\'evy stable
and Mittag-Leffler distributions with little programming and computational efforts and with the same accuracy of these specialized packages. Furthermore,
from the Mikusinski's  representation one will be able to derive some asymptotic expressions with special relevance to physical applications.

\section{The algorithm}

The Mikusinski's integral representation (\ref{Miku}) involves a
simple proper  integral
of a smooth function on the interval $\varphi\in[0,\pi]$. Furthermore,
it is easy to obtain   from Eq. (\ref{Miku}) some
analogous formulas for the
derivatives of $g_\alpha(x)$ and its cumulative distribution function
$\Lambda_\alpha(x)$. In particular, we have
\begin{eqnarray}
\label{cdf}
\Lambda_\alpha(x) &=&  \frac{1}{\pi}
\int_0^\pi  e^{-u }\,d\varphi, \\
\label{gl}
g'_\alpha(x) &=&  \left(\frac{\alpha}{1-\alpha}\right)^2 \frac{1}{\pi x^2}
\int_0^\pi u^2e^{-u }\,d\varphi   -
 \frac{1}{ 1-\alpha } \frac{g_\alpha(x) } {x}, \nonumber \\
\end{eqnarray}
and
\begin{eqnarray}
\label{gll}
g''_\alpha(x) &=&  \left(\frac{\alpha}{1-\alpha}\right)^3 \frac{1}{\pi x^3}
\int_0^\pi u^3e^{-u }\,d\varphi \nonumber   \\
 &   & -   \frac{3}{1-\alpha}
 \frac{g'_\alpha(x)}{x}  -
 \frac{1+\alpha}{ (1-\alpha)^2}\frac{g_\alpha(x)}{ x^2},
\end{eqnarray}
where $u$ is given by Eq. (\ref{u}). The formulas for the Mittag-Leffler case can be obtained directly
from Eqs. (\ref{feller}) and (\ref{feller1}).
Fig. 1 depicts the
integrand in Eq. (\ref{Miku}) for some typical values of $\alpha$ and $x$.
\begin{figure}[t]
\includegraphics[width=1\linewidth]{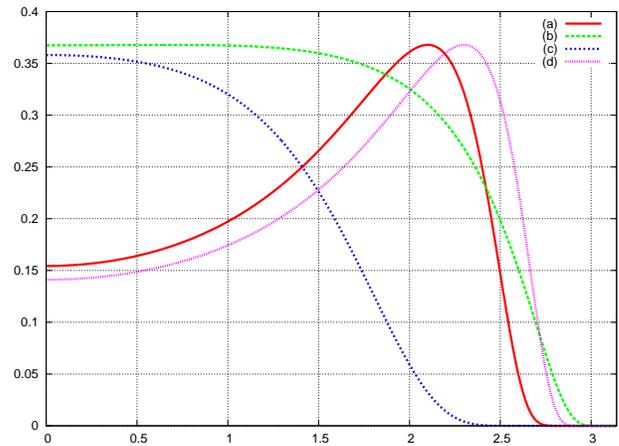}
\caption{The integrand $f=u e^{-u}$ of the Mikusinski's representation (\ref{Miku})
for some values of $\alpha$ and $x$ in the
interval $\varphi\in[0,\pi]$. The curves (a)-(d) correspond,
respectively, to the following values of $(\alpha,x)$:
$(0.6,1.0)$, $(0.2,0.1)$, $(0.5,0.2)$, and $(0.5,1.5)$.}
\label{fig1}
\end{figure}
The integrands for the derivatives of  $g_\alpha(x)$ and for the
 Mittag-Leffler case have similar aspects. They
are all of the type $f_n = u^ne^{-u}$ on the interval
 $\varphi\in[0,\pi]$, where $u$ is given by Eq. (\ref{u}).
It is easy to show that: $f_n = u_0^ne^{-u_0}$ and $f'_n = 0$ for $\varphi = 0$, with $u_0=(1-\alpha)(\alpha/x)^{\alpha/(1-\alpha)}$;
  $f_n\to 0 $ for $\varphi\to\pi$; and that $f_n$ has a  maximum for
  $\varphi$ such that $u=n$, provided $u_0<n$.
   Although the position of such maximum does
  depend on $\alpha$ and $x$, its value ($f_n=n^ne^{-n}$) depends only on $n$. For a given $\alpha$, larger values of $x$ displace the maximum
  towards $\varphi = \pi$, while smaller values does towards $\varphi=0$.
   If $u_0\ge n$, the unique maximum of $f_n$ is at $\varphi=0$.
For the cumulative distribution function (\ref{cdf}), the integrand corresponds to $n=0$. In particular its maximum is located at $\varphi=0$, with  $f_0=1$, irrespective of the values of $x>0$ and $0<\alpha<1$.
 All these functions are well-behaved on the interval $\varphi\in[0,\pi]$
and, consequently,  integrals like Eqs. (\ref{Miku}), (\ref{cdf}), (\ref{gl}), and (\ref{gll})  can be evaluated  numerically without major problems.

\begin{figure}[tb]
\includegraphics[width=1\linewidth]{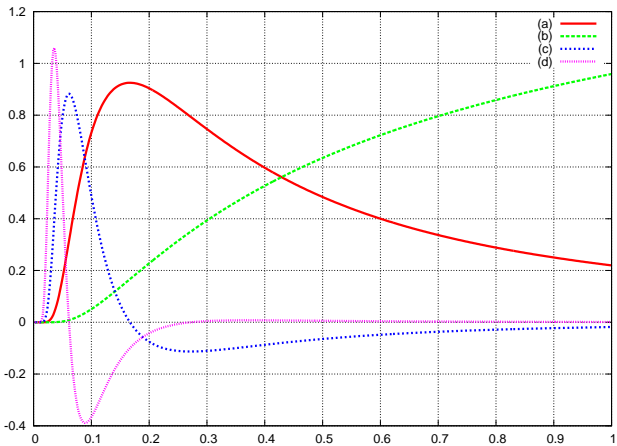}
\caption{Plots for $g_{1/2}(x)$ (a), $2\Lambda_{1/2}(x)$ (b), $(1/15)g'_{1/2}(x) $ (c), and $(1/500)g''_{1/2}(x)$ (d),
for $0\le x \le 1$, calculated by the numerical integration of Eqs. (\ref{Miku}), (\ref{cdf}), (\ref{gl}), and (\ref{gll}) by means of the SLATEC \cite{slatec}  adaptative integration routine {\tt DQAGS}.
The relative errors for all the four curves, calculated with respect to the exact Smirnov's
distribution (\ref{Smirnov}) in the interval $0\le x \le 5$, are smaller
than $3\times  10^{-8}$.
 (Numerical code available at
\cite{html}.)}
\label{fig2}
\end{figure}

We have set up an
adaptive integration scheme  based
on the publicly available {\tt DQAGS} routine of SLATEC \cite{slatec}.
We could integrate Eqs. (\ref{Miku}), (\ref{cdf}), (\ref{gl}), and (\ref{gll}) with little
computational effort demanding a relative error in {\tt DQAGS} smaller than $10^{-8}$, which is typically attained with about 10 iterations of
the global adaptative scheme of the routine.  Our FORTRAN code, available at \cite{html}, has demonstrated
to be extremely robust and reliable.
 In order to test it,
 we have used the case corresponding to the L\'evy distribution with $\alpha=1/2$, for which an explicit form for
the probability density function  is known, namely the so-called Smirnov's distribution
\beq
\label{Smirnov}
S(x) = g_{1/2}(x) = \frac{e^{-1/4x}}{2\sqrt{\pi}x^{3/2}}.
\eeq
Fig. \ref{fig2}
shows the functions $g_{1/2}(x)$, $\Lambda_{1/2}(x)$, $g'_{1/2}(x)$, and $g''_{1/2}(x)$ evaluated  numerically with our code. As we see, we can calculate
$g_\alpha(x)$ with very good accuracy and in an efficient way. The corresponding data (500 points for each curve) for plots like those ones depicted in Fig. \ref{fig2} are generated instantaneously in a Intel Core i7 computer. The relation (\ref{feller1}) and the
Smirnov's distribution (\ref{Smirnov}) allow us to test also the
Mittag-Leffler case  since they imply that
\beq
\label{ML12}
\rho_{1/2}(x) = \frac{e^{-x^2/4}}{\sqrt{\pi}}.
\eeq
Our numerical procedure works with similar accuracy for this particular Mittag-Leffler probability density, with the usual caveats related to
extremely small values of $x$ in (\ref{ML12}), which correspond to large values of $x$ in $g_{1/2}(x)$ according to (\ref{feller1}).
We also checked the good accuracy of our algorithm by comparing the
 output with Nolan's {\tt STABLE} package \cite{Ref}.

The numerical evaluation of the probability densities for
extreme values of $x$ and $\alpha$ is quite delicate due to
convergence and roundoff problems. For a fixed $\alpha$ and $x\to 0$
and $x\to \infty$,  the asymptotic formulas (\ref{asymp1}),  (\ref{asymp2}), (\ref{asympML1}), and
 (\ref{asympML2}) can be indeed used to estimate the probability densities.
For fixed $x$ and $\alpha$ very close to $0$ and $1$, other asymptotic
expressions are necessary. For small
$\alpha$, we have from Eq. (\ref{u})
\beq
\label{u1}
u \approx 1+
\alpha\ln\alpha+
\left( \ln\frac{\varphi}{x\sin\varphi}- \frac{\varphi}{\tan\varphi}
\right)\alpha.
\eeq
For a fixed  $x\in(\alpha,1/\alpha)$, we have  $u \approx 1+
\alpha\ln\alpha$ for small enough $\alpha$
 and $0\le \varphi < \pi$, leading to
$ue^{-u}\approx e^{-1}\left(1 -(\alpha\ln\alpha)^2/2\right)$ for small $\alpha$ and $0\le \varphi < \pi$.
Applying  these results in Eq. (\ref{Miku}), one
has
\beq
\label{asymp3}
g_\alpha(x) \approx \frac{\alpha}{ex},
\eeq
valid
for small $\alpha$ and $\alpha < x < 1/\alpha$. For $x<\alpha$, the
approximation (\ref{asymp1}) is still valid for small $\alpha$.
In particular, we always have $g_\alpha(x)\to 0$ for $x\to 0$,
irrespective of the value for $\alpha$. Since the hypothesis of
$\alpha < x < 1/\alpha$ was explicitly used, the approximation (\ref{asymp3}) is
not supposed to be accurate for   $x\to\infty$. In this case, Eq. (\ref{asymp2})
is the correct asymptotic expression for  $g_\alpha(x)$.
For $\alpha$ close to $1$, the situation is a little bit more involved.
Introducing $1-\alpha = \varepsilon > 0$,
we have from Eq. (\ref{u})
\beq
\label{u2}
u \approx \frac{\varepsilon}{x^{1/\varepsilon}} \frac{\varphi}{\sin\varphi}e^{-\varphi/\tan\varphi},
\eeq
valid for $\varepsilon\approx 0$,  $0 < x < \infty$,
and $0\le \varphi < \pi$.
For $x>1$, $u\to 0$ for small $\varepsilon$, implying that
$g_{1-\varepsilon}(x)\to 0$. For $x<1$, $u\to \infty$,
also implying  $g_{1-\varepsilon}(x)\to 0$.
 Since, according to Eq. (\ref{laplace}),
$g_\alpha(x)$ is supposed to be normalized for any
value  of $\alpha$, $g_\alpha(x)\to\infty$
for  $x\to 1$ and $\alpha\to 1$. Hence, for $\alpha$ close to $1$,
$g_\alpha(x)$ should be strongly peaked around $x=1$, resembling an approximation
for a $\delta$-function. Such behavior could also be inferred by
considering the limit $\alpha\to 1$ directly in Eq. (\ref{laplace}).

\subsection{The maxima of the distributions}

\begin{figure}[tb]
\includegraphics[width=1\linewidth]{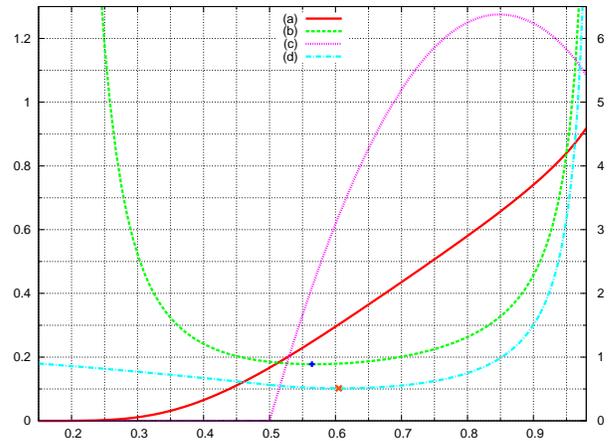}
\caption{Location and value of the maxima of the probability density
functions of L\'evy (curves (a) and (b)) and Mittag-Leffler (curves (c) and (d)) distributions. Curves (a) and (c) correspond to the location   $x_*(\alpha)$ of the maxima (left scale), as function of $\alpha\in(0,1)$ (horizontal axis). Curves (b) and (d) are the value of the maxima (right scale) as function of $\alpha$. The marked points correspond to
 the shortest maxima
for each distribution, $(\alpha = 0.567,g_\alpha(x_*)=0.888)$ and $(\alpha = 0.605,\rho_\alpha(x_*)=0.509)$.}
\label{fig3}
\end{figure}

As an application of our numerical procedures,
we will explore some properties of
the distribution $g_\alpha(x)$ and $\rho_\alpha(x)$.
The location of the maxima of these probability density functions is certainly
 pertinent to the understanding of the statistical
  processes governed by
 them.
Let us consider fist the case of
$g_\alpha(x)$.
 The condition determining the location $x_*(\alpha)$ of the maximum
 of the probability density is, of course,
$g'_\alpha(x_*)=0.$ From the approximations discussed in
the preceding section, we have
that $0 < x_*(\alpha) < 1$ for $0<\alpha<1$ and that $g_\alpha(x_*)\to\infty$ for $\alpha\to 0$ and for $\alpha\to 1$.
The zero of $g'_\alpha(x)$ can be localized in the interval $(0,1)$ with a
prescribed accuracy by using, for instance, a simple bisection method. Since
we have a procedure to calculate $g''_\alpha(x)$, one could even implement
a refinement for the determination of $x_*(\alpha)$
based, for instance, in Newton-Rapson method.
Fig. \ref{fig3} shows the values of $x_*(\alpha)$ and $g_\alpha(x_*)$
for $0<\alpha<1$.
Notice that, as expected,
  we have
that $g_\alpha(x_*)\to\infty$ for $\alpha\to 0$ and $\alpha\to 1$,
 in agreement with the approximations of last section. The minimal value of $g_\alpha(x_*)$ is attained when $\alpha = 0.567$,
corresponding to the    one-sided $\alpha$-stable
L\'evy distribution with shortest maximum, for which $g_\alpha(x_*)\approx 0.888$.
\begin{figure}[tb]
\includegraphics[width=1\linewidth]{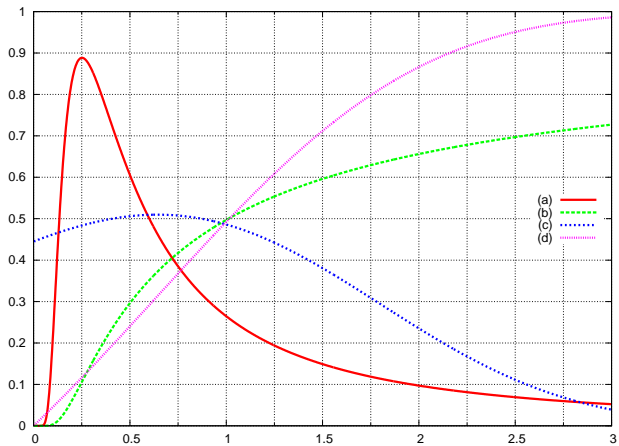}
\caption{L\'evy and Mittag-Leffler probability densities  with shortest maxima (respectively, curves (a), with $\alpha = 0.567$, and (c), with $\alpha = 0.605$, see Fig. \ref{fig3}) and their respective cumulative distribution function (curves (b) and (d)) in the interval $0<x<3$.}
\label{fig4}
\end{figure}
(Fig. \ref{fig4}).

The situation for $\rho_\alpha(x)$ is rather
 more involved. For $\alpha\approx 1$,
$\rho_\alpha(x)$ is similar to $g_\alpha(x)$, both resembling approximations
of a $\delta$-function around $x=1$. Such behavior for the
 Mittag-Leffler case can also be inferred
directly from the definition (\ref{laplace1}), by considering the limit
$\alpha\to 1$. However, in contrast with the previous case, for $\alpha\approx 0$, $\rho_\alpha(x)\approx e^{-x}$, as one can also see by evaluating the limit
$\alpha\to 0$ in Eq. (\ref{laplace1}). Hence, for small $\alpha$, the maximum of $\rho_\alpha(x)$  located at $x=0$ and is given by  $\rho_\alpha(0)\approx 1$. In fact, we could verify numerically that for $\alpha < 1/2$, the maximum of $\rho_\alpha(x)$ is is always at $x=0$ and is given by Eq. (\ref{asympML1}). For $\alpha > 1/2$, we have $\rho'_\alpha(0)>0$ and  the function $\rho_\alpha(x)$ attains a maximum for $x>0$ and then decays. Curiously, as $\alpha$ increases, $x_*(\alpha)$ also increases and even exceed $x=1$, and then
return to $x=1$, but from the right-handed side. This behavior, which will be crucial for the discussion of the next section, is depicted by the
curve (c) in Fig. \ref{fig3}. The L\'evy and Mittag-Leffler distribution
with shortest maxima are plotted in Fig. \ref{fig4}.

\section{Distribution of Lyapunov exponents in intermittent systems}

 We can also apply our numerical procedures to elucidate
 some dynamical problems of physical interest. This is the case,
for instance,
 of the distribution of Lyapunov exponents in intermittent systems such as
  the Pomeau-Manneville maps  $x_{t+1}=x_{t}+ax_{t}^{z}\, (\mbox{mod}\, 1)$
considered recently in \cite{KB}. For $z>2$, theses systems are
known to exhibit, for nearby trajectories, a  subexponential
 deviation of the type $\delta x_{t}\sim\delta x_{0}\exp(\lambda_{\alpha}t^{\alpha})$, where $\alpha=1/(z-1)$.
According to the Aaronson-Darling-Kac (ADK) theorem \cite{ADK}, for randomly
distributed
initial conditions and sufficiently large times, the ratio
$\lambda_{\alpha}/\left\langle \lambda\right\rangle$, where $\left\langle \lambda\right\rangle$ is a suitable   average for the exponents $\lambda_{\alpha}$, converges in distribution terms towards a Mittag-Leffler random variable
with unit first moment and index $\alpha\in (0,1)$.
Such statistics was also considered previously in Ref. \cite{AA} from the numerical point of view.
Some recent numerical
works \cite{KB} have reported a regular tendency of $\lambda_{\alpha}$ be
smaller than the average $\left\langle \lambda\right\rangle$ for large values of $z$ (small $\alpha$).
In fact, in \cite{KB} the first moment $\left\langle\lambda\right\rangle$ is calculated differently from the ADK theorem, it is obtained there from a continuous-time stochastic model, but its values are, for the considered Pomeau-Manneville maps, the same of the ADK ones.
Since $\lambda_{\alpha}/\left\langle \lambda\right\rangle$ is a random Mittag-Leffler variable, we can evaluate
the probability of having $\lambda_{\alpha}<\left\langle \lambda\right\rangle$
\begin{equation}
\label{prob}
\mbox{Prob}(\lambda_{\alpha}<\left\langle \lambda\right\rangle)=\int_{0}^{1}\rho^{(r)}_{\alpha}(x)dx=1-\Lambda_{\alpha}(\Gamma^{1/\alpha}(1+\alpha)),
\end{equation}
where $r=\Gamma^{1/\alpha}(1+\alpha)$ assures that $\rho^{(r)}_{\alpha}(x)$
has unit fist moment, as required by the ADK theorem.
 \begin{figure}[tb]
\includegraphics[width=1\linewidth]{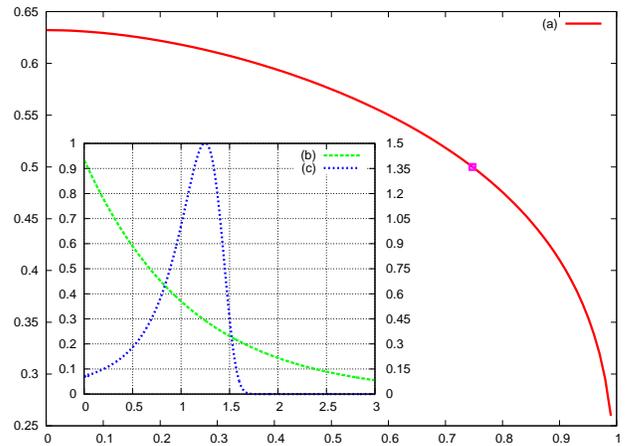}
\caption{Curve (a): probability of having $\lambda_\alpha < \left\langle \lambda\right\rangle$ in Pomeau-Manneville maps as function of
$\alpha\in(0,1)$, according to Eq. (\ref{prob}).
The marked point ($\alpha\approx 0.747$) corresponds to the equiprobability.
Detail: curves (b) and (c) are, respectively,
the probability density of Mittag-Leffler distribution with unit first
moment and $\alpha=0.1$ (left $y$-scale) and $\alpha=0.9$ (right $y$-scale)
in the interval $0<x<3$. It is clear why the probability decays for increasing $\alpha$: the
probability density tends towards a $\delta$-function centered in $x=1$, but
from the right handed side. (See curve (c) in Fig. \ref{fig3}).
}
\label{fig5}
\end{figure}
Fig. \ref{fig5} depicts
this probability as function of $\alpha$. For $\alpha\to 0$, we have $\mbox{Prob}(\lambda_{\alpha}<\left\langle \lambda\right\rangle)\to 1-1/e\approx 63 \%$. As we see, the tendency reported in \cite{KB} of having $\lambda_\alpha < \left\langle \lambda\right\rangle$ can be clearly understood from the ADK theorem. Moreover,
the aspect of the Mittag-Leffler distributions for small $\alpha$ and
$\alpha\to 1$ explains  why these intermittent systems do exhibit such
kind of
behavior. For $\alpha\to 0$, the probability density function of the
Mittag-Leffler distribution has the form $\rho_\alpha(x)\approx e^{-x}$ (See Fig. \ref{fig5}). Its maximum is located at $x=0$, and it is clear that
the typical values of the random variable are always
 smaller than its average.
On the other hand,  for $\alpha\to 1$ ($z\to 2^+$), the probability density resembles
a $\delta$-function with center approaching $x=1$, but from the right handed sided, see Fig. \ref{fig5}. In this case, the typical values of the random variable
remain close the value of its average. Also from figure, we have that for $\alpha > 3/4$ the probability of having  $\lambda_{\alpha}>\left\langle \lambda\right\rangle$ is favorable over Eq. (\ref{prob}). (In fact, the   equiprobability corresponds to
 $\alpha\approx 0.747$.)
In terms of the distribution of Lyapunov exponents for
  the Pomeau-Manneville maps, this would correspond to have a
  slight
  predominance
  of   $\lambda_{\alpha}$ greater than the average $\left\langle \lambda\right\rangle$ for small $z>2$. This seems, in fact, marginally
  evident from \cite{KB}, but further work is necessary to
  establish this fact with the same certainty of the behavior for large $z$.
This kind of problem  in intermittent systems are very interesting and certainly deserve a deeper investigation.

\section{Final remarks}

Motivated by the closed form formulas in terms of the Meijer G-functions
for the probability densities $g_\alpha(x)$ of one-sided L\'evy distributions with rational $\alpha=l/k$ proposed by Penson and  G\'orska in
\cite{PG},
we have introduced a numerical scheme  for the computation of $g_\alpha(x)$
for any real $\alpha\in (0,1)$. By exploring the relation between
one-sided L\'evy and Mittag-Leffler distributions, we extend our procedures to include the evaluation  of the
the probability densities $\rho_\alpha(x)$ of Mittag-Leffler
distributions. The main advantage of our numerical
approximation is that it can be applied for any value $\alpha$, while
Penson and  G\'orska formula (\ref{meijer}) is rather problematic for fractions with large denominators. As an application of our procedures, we
determine the maximum location and value for the densities $g_\alpha(x)$
and $\rho_\alpha(x)$ as function of the index $\alpha\in (0,1)$. We show   that $\alpha \approx 0.567$ and
 $\alpha \approx 0.605$   correspond, respectively,
  to the   one-sided
L\'evy and Mittag-Leffler distributions with shortest maxima. Furthermore,
we use our numerical procedure for the evaluation of Mittag-Leffler distribution to show that a recently described statistical behavior for intermittent systems \cite{KB}, namely the predominance of having Lyapunov exponents
$\lambda_\alpha$ smaller than the theoretical average $\langle \lambda\rangle$ for Pomeau-Manneville maps with large $z$,
is nothing else than a consequence of the Mittag-Leffler statistics.
We hope our numerical procedures could
be useful for this kind of study.

\acknowledgements

This work was supported by FAPESP and CNPq.


\begin{references}
\bibitem{procedure} For numerical evaluations, it is more convenient
to eliminate the command {\tt convert} from the Maple procedure presented in \cite{PG}
\end{references}
\end{document}